\documentstyle[preprint,aps]{revtex}
\begin{document}

\draft
\flushbottom

\title{The Puzzling Stability of Monatomic Gold Wires}

\author{J. A. Torres$^1$, E. Tosatti$^{1,2,3}$, A. Dal Corso$^{2,3}$, F. Ercolessi$^{2,3}$,\\
J. J. Kohanoff$^1$, F. D. Di Tolla$^{2,3}$, and J. M. Soler$^4$\\ }

\address{
1: Abdus Salam International Centre for Theoretical Physics (ICTP),
Trieste, Italy.\\
2: International School for Advances Studies (SISSA), Trieste, Italy.\\
3: Istituto Nazionale di Fisica della Materia (INFM), Trieste, Italy. \\
4: Dep. F\'{\i}sica de la Materia Condensada. Univ. Aut\'onoma de Madrid,
Spain.
}

\date{22 December 1998}

\maketitle

\begin{abstract}
We have examined theoretically the spontaneous thinning process 
of tip-suspended nanowires, and subsequently studied the structure 
and stability of the monatomic gold wires recently observed by 
Transmission Electron Microscopy (TEM).
The methods used include thermodynamics, classical many-body force 
simulations, Local Density
(LDA) and Generalized Gradient (GGA) electronic structure calculations 
as well as ab-initio simulations including
the two tips. The wire thinning is well explained in terms of a 
thermodynamic tip suction driving migration of surface atoms from
the wire to the tips. For the same reason the monatomic wire becomes 
progressively stretched. Surprisingly, however, all calculations 
so far indicate that the stretched monatomic gold wire should be 
unstable against breaking, contrary to the apparent experimental 
stability. The possible reasons for the observed stability are discussed.
\end{abstract}


\tightenlines 
In the last few years the properties of atomic size 
metallic contacts have attracted great attention both for 
fundamental and for practical reasons. 
Experimental formation of these contacts has been achieved 
using different techniques and in diverse conditions ranging 
from  room temperature under
atmospheric pressure \cite{PascPRL}, to  low temperatures in 
ultrahigh vacuum \cite{Ruit}.
The relationship between
mechanical structure and electrical properties has been studied 
by simultaneously measuring load force and  conductance  \cite{Rubio}.
On the theoretical side, the mechanical structure
and evolution of a tip-surface mechanical contact 
has been modelled using classical molecular 
dynamics  (MD) simulations with many body potentials \cite{Tomagnini,gneck,PascPRL}, 
and also by ab-initio MD simulations.\cite{Barnett}.
The conductance of these necks has been calculated using free-electron
\cite{PRB1}, tight binding \cite{Alvaro} and ab-intio schemes \cite{Lang,Ditolla}.
The relationship between mechanical load and conductance has been studied using
different models like free electron \cite{PRL}, 
tight binding \cite{Todorov} and LDA \cite{Dani}. 

Very long tip-suspended nanowires of surprising regularity and stability
were found in simulations close to the melting 
point \cite{Tomagnini,tomagnini}, and
subsequently confirmed experimentally \cite{kuipers}.
Questions about the structural and thermal behavior of ideally infinite
nanowires have recently stirred  an upsurge
of theoretical  \cite{Gulseren1,Gulseren2} and experimental interest 
\cite{Ohnishi}. New observed structural features of long nanowires
include a striking regularity 
\cite{Ohnishi}. Both crystalline and noncrystalline morphologies 
have been predicted \cite{Gulseren2} and observed \cite{Ohnishi}. 
Very recently the exceptional ductility and diffusivity
of metallic gold has been put to use in creating the ultimately 
thin nanowire, consisting of just a single atomic strand 
\cite{Taka,Yanson,Ohnishi}. Direct TEM imaging,\cite{Taka} in
particular, has shown that a thin multi-shell wire suspended
between two (110) tips will gradually and spontaneously thin down to smaller
and smaller radii, until the two tips are eventually connected by 
just a 4-atom long monatomic wire. Two of
these four Au atoms stretch out of a well-tapered tips, the remaining 
two are freely suspended in
vacuum in an astonishingly stretched configuration. The  stretching magnitude 
increases with time,
the average atomic distances growing from about 3 to about 4 \AA, yet the
wire is not seen to break for many seconds. In different measurements
\cite{Yanson,Ohnishi}, the wire was not directly imaged, but conductance 
plateaus close to and
below $G_0 = 2e^2/h$ indicated that the tip-suspended wire must 
have been monatomic and conducting. For that case, however, no 
structural information is available. These findings
raise several questions, among them: 

$i$) why are the wires thinning down with time, and what is the
origin of the stretching force
which appears to act on the monatomic wire even in the absence of
any mechanical tip withdrawal? 

$ii)$ how can we understand the structural stability
of such an exceptionally stretched monatomic wire, particularly 
in terms of its electronic structure?

$iii$) how can we explain the wire conductance?

While we have not yet addressed $iii$, the
scope of this note is to point out that thermodynamic reasoning
readily explains point $i$, whereas preliminary calculations and 
simulations performed with state-of-the-art methods totally 
fail to account for the wire stability issue of point $ii$. This is all 
the more surprising since these methods, tuned on the
properties of atomic Au, are known to work quite well for bulk 
and surfaces of metallic Au \cite{ref7}. In the following we will 
address first the thinning. Secondly we will describe
our attempts at understanding cohesion and breaking of the infinite
wire and of the tip-suspended short wire, and finally 
speculate on what might be the explanation for the unusual stability.

\section{Why do suspended wires thin down, and why is the monatomic wire
stretched?}

All wires suspended between two tips appear to thin down steadily
with time, on a scale which is visible within seconds.\cite{Ohnishi,Taka}.
 The rate of decrease moreover appears somewhat
faster for thinner wires. Once the wire is monatomic, it
becomes increasingly stretched with time, even though the tips are fixed
relative to one another. 

We propose the following simple thermodynamical theory.
For a sufficiently thick  wire of radius $R$, the free energy per atom is:
\begin{equation}
\varepsilon (R) = \varepsilon _0 + 2 \gamma / R \rho 
\label{eq1}
\end{equation}
where $\varepsilon _0$ is the bulk free energy (essentially identical to
the cohesive energy), $\gamma$ is the surface free energy
per unit area and $\rho$ is the bulk density. Taking one atom 
from the wire to the tip, the
free energy decreases, and in turn this represents an effective
suction force exerted by the tips on the wire atoms. In presence of surface
diffusion, the wire surface atoms will drift towards the tips, and
the wire radius will thus decrease. For a surface diffusion 
coefficient $D_{s}$ and an Einstein-related 
surface mobility $\mu_{s} = D_s / k_B T$,
an atom on the wire surface will drift with velocity $v$ towards the tip, 
and will be absorbed by it in an average time $t \approx L/4v$, where
L is the wire length. Due to that, the wire radius will
decrease roughly by $\delta R = - 1/ 2 \pi R L \rho$. The free energy
will first of all decrease by $-2  \gamma / R \rho$ since an atom 
has left the wire for the (bulk-like) tip. However the wire radius 
decrease implies an additional free energy cost, all other
atoms being left  behind in a now thinner wire, amounting to 
$-(2 \gamma / R^2 \rho )( \pi R^2 L \rho - 1 )\delta R$. 
This cancels one half of the previous free energy decrease
(neglecting ${1\over N}$ terms), so that the 
total free energy
change for an atom leaving the wire and going to the tips is, 
to lowest order in ${1 \over R}$, 
\begin{equation}
\Delta E = -\gamma / R \rho
\label{eq2}
\end{equation}
Now the average velocity of surface atom, drifting from the wire
middle towards the closest tip, approximately a distance $L/4$ away,
will be $v=\mu_{s}(\Delta E / \Delta x)$, or 
$v=4 \gamma \mu_{s} / R L \rho $. 
Assuming all surface particles within an atomic diameter 
$2r_0$, whose number is $N_s =
4 \pi L \rho r_0 (R - r_0)$, to drift in this manner,  
the radius will decrease with a thinning rate:
\begin{equation}
\frac{\Delta R}{\Delta t} = \frac{\delta R N_s}{L/4v} 
= 128 \pi \frac{\gamma \mu_s r_0^4 (1 - r_0/R)}{3 R L^2}
\label{eq4}
\end{equation}
where $\rho = 3/4 \pi r_0^3$ has been used.
The predicted wire thinning rate calculated
from this formula for wires of fixed length and variable 
radius, with a surface free energy 
$\gamma_{Au}\sim 1.3 J/m^2 $ and diffusion constant $D_{s}$ 
ranging from $10^{-5}$ (typical of
liquid Au) to $10^{-7}$ cm$^2$/s, is shown in Fig. 1. As can be seen a
typical value for the thinning rate can be in the order of 1 \AA/s for
a nanometer thin wire, which seems of the correct magnitude 
to explain the experimentally observed thinning at room 
temperature \cite{Taka}.
The thinner and shorter a wire is, 
the faster it will further thin down, until it gets monatomic. 
This behavior should be  amenable to future experimental verification.

Once the thinning has reduced it to monatomic,
the wire can only stretch under the tip suction, until it will eventually 
break. It is therefore mandatory to estimate the stability of a 
wire under stretch.

\section{The infinite monatomic gold wire}

The electronic structure of an infinite, regular gold monatomic 
wire has been calculated as a function of the interatomic spacing. The
calculation was performed within  density functional theory
in the LDA, and also in GGA, using 
the nonlocal ultrasoft pseudopotentials described in reference
\cite{Corso}. A repeated cell geometry was used, with a large 
inter-wire distance of 10.6 \AA, and a plane wave expansion 
with a cutoff energy of 25 Ry for the wavefunctions and 200 Ry for
the charge density.
The calculated total energy curves versus distance are shown in 
Fig. \ref{fig2}. They have minima of about 2.2 eV/atom (LDA), or
1.7 eV/atom (GGA)~\cite{PW91} at an 
equilibrium distance of about 2.50 \AA\ (LDA) or 2.55 \AA\ (GGA). 
For reference, with the same pseudopotentials and cut-offs, 
the calculated cohesive energy for bulk Au is 4.4 eV (LDA),   
3.3 eV (GGA), against an experimental value of 3.9 eV. 
For comparison we also present in Fig. \ref{fig1} the wire
cohesive energy calculated using an alternative local orbital 
based LDA and GGA scheme, the SIESTA program \cite{siesta,siestacal}, 
as well as those given by the empirical glue potential
for gold \cite{glue}. The wire cohesive energy per atom is 2.8 eV (LDA),
2.1 eV (GGA),~\cite{PBE}  and only 0.4 eV (glue). The corresponding 
bulk cohesive energies
provided by the local orbital calculation is 5.4 eV (LDA)
3.9 eV (GGA), 3.9 eV (glue). The four ab-initio calculations are 
in substantial agreement (we have not attempted a further improvement
by extending the basis) while the glue model is seen to severely
underbind, and should be discarded for such an 
ultra-thin wire.~\cite{note1glue} 
Stretching the wire interatomic distance from its
equilibrium value to, say, 3.5 \AA\ would cost, according to 
the local orbital GGA calculation, a stretching energy of 
about $4.0\pm 0.5$ eV, 
nearly exactly compensating the energy gain corresponing to 
one gold atom sucked by one of the tips, namely
the bulk cohesive energy, which is 3.9 eV. This would seem a very
reasonable balance, supporting the concept that tip suction 
is indeed responsible for the observed wire stretching. 

However, this reasoning is not further supported by the electronic
structure calculations where we find that even a relatively modest 
stretching endangers the wire stability. Using for simplicity a four 
atom cell with periodic boundary conditions, we
tested for longitudinal stability by calculating total energy 
changes when one interatomic spacing is extended from $a$ to $a'$, 
while forcing the three remaining interatomic distances to shrink
so as to keep a total lenght of $4a$ (see Fig. \ref{fig3}). 
The energy increase obtained by increasing $a'$ relative to $a$ 
represents the energy barrier which
prevents the wire from longitudinal breaking. To our surprise, 
we found (Fig. 3) that the barrier is positive  at the equilibrium 
spacing of 2.55 \AA~ and just above, but disappears very
quickly when the wire is stretched, making it unstable 
below 3 \AA.~\cite{note2} 
This result is puzzling, in view of the exceptional stability 
observed experimentally. However, there
is still a distinct possibility that the real 4-atom wire hanging between 
well-tapered tips could in many ways behave differently from the
tip-less wire just studied. We thus decided to investigate further 
a more realistic case.

\section{Short monatomic wire between (110) tips: ab-initio simulation}

For the first-principles MD relaxation, we generated a starting configuration
in the following manner. We started with an FCC bulk-ordered 
cubic-shaped piece of 240 atoms. Keeping both extreme layers rigid,  
we separated them at constant velocity 
along the (110) direction and simulated the evolution of the
neck using glue model MD (so long as atomic coordination not too poor,
the glue model provides quite reasonable interatomic forces for Au.)  
The stretching velocity was  $\approx$  0.14 m\AA/fs, and
the temperature was kept very low. A neck formed and broke. After 
neck breaking we stopped the simulation and inserted three 
additional atoms in a row, forming a wire bridging 
the opened gap. Finally, we relaxed the new configuration, and the 
interatomic spacing of the relaxed wire reached about 2.9 \AA. 
We further pulled the tips, stretching the
monatomic wire up to the  very moment before
it would break and stopped there, finally relaxing and checking for
stability with glue forces. The
33 central atoms of this last configuration, shown in Fig.~\ref{fig3}
form our initial  configuration, with two tapered tips and 
a 4-atom wire with an interatomic spacing  of about 3.5 \AA~.

Starting with this configuration, we calculated the electronic 
structure and the forces on all 33 atoms
using the local basis LDA \cite{siestacal2}. 
With these forces we conducted a short ab initio MD
relaxation of 40 steps of 5 fs each plus 60 steps of
7 fs each. The resulting ab-initio evolution after 100,
200, 340, 480 and 620 fs is also shown in Fig. \ref{fig3}. 
We see that interatomic distances in the wire change rapidly.
However, while most of them shorten, especially close to the tip, one
distance is growing with time, and the wire is clearly 
breaking. It appears therefore that within
this approximation, the short wire hanging from the tips is unstable, 
very much as the infinite wire was.

\section{Discussion}
We believe we have understood the thinning and stretching 
of wires between tips, and provided a thinning rate dependence 
upon the wire length and radius that can be checked experimentally. 
The exceptional stability of the stretched monatomic wire 
remains instead mysterious since
state-of-the-art calculations predict that it should break 
by far. While independent
measurements would clearly be desirable in order to confirm the 
large spacings observed
in reference \cite{Taka}, we can presently only speculate on 
the possible reasons why our theory
might be in error on this problem. 

A first possibility is that 
electron correlations, weak in bulk Au, could become strong in 
severely stretched wires, owing to the
exceedingly poor coordination, and to the large interparticle 
distance. In particular, typical one
dimensional instabilities, involving antiferromagnetism or dimerization
could play a role. Of them, dimerization is allowed in our 
simulation, but magnetism is not. While this is clearly an
avenue to be explored, it seems at this stage unlikely that it 
could alter the energetics very drastically. All the same,
if the stretched wire turned magnetic (and presumably also insulating),
the corresponding small energy decrease might be sufficient
to generate a small change of curvature and therefore
a small barrier against breaking.

A second factor which we neglected throughout is long-range
correlations, and particularly van der Waals forces. Van der Waals 
attraction is known to be strong especially between blunt tips 
\cite{Tosa92}. 
This attraction
could be enhanced by the presence of a conducting wire joining the two
tips, thus providing the missing barrier. This effect is absent in LDA/GGA
which does not include long-range dispersion forces.

An alternative possibility is that the TEM electron beam could 
be influencing, by heating or charging, the wire, and causing 
its anomalous behavior. However both tips are electrically
grounded, and it is further estimated \cite{Takapri}
that only about one electron per second collides elastically 
with the wire, the inelastic cross section being even smaller.

Clearly the problem remains open, and more experimental and 
theoretical work will be needed to solve it.

We are grateful to Prof. K. Takayanagi and his group for enlightening
discussions.
We acknowledge financial support from INFM/G (Project "Nanowires"),
INFM/F (PRA LOTUS), by MURST through COFIN97, and 
Spain's DGES, project PB95-0202.
Work by J.A.T. was carried out under TMR grant ERBFMBICT972563.

\begin{figure}
\narrowtext
\begin{center}  
\end{center}

\caption{ Thinning rate of tip suspended 20 \AA\ long
gold wires, from Eq. 2, calculated for different surface diffusion
constants. }
\label{figdRdt}
\end{figure}

\begin{figure}
\narrowtext
\begin{center}
\end{center}

\caption{Cohesive energy per atom for an infinite monatomic Au wire.}
\label{fig1}
\end{figure}

\begin{figure}
\narrowtext
\begin{center}
\end{center}

\caption{ Energy variation upon attempted wire breaking. The wire
is not stable when the interatomic spacing is stretched above ~2.8 \AA . 
}
\label{fig2}
\end{figure}

\begin{figure}
\narrowtext
\begin{center}
\end{center}
\caption{ Snapshots of the ab-initio MD relaxation for increasing time.
 The spontaneous breaking of the wire is evident.} 
\label{fig3}
\end{figure}



\begin{references}
\bibitem{note1glue}
There is in reality a deeper minimum (1.2 eV) for the glue model
at 1.73 \AA, but it is clearly unphysical.
\bibitem{note2}
A reversal of concavity, from positive to negative in the energy
vs. distance curve of Fig.~2 occurs in fact at $\sim 2.8$ \AA. 
It can be seen as
giving rise to the instability, e.g. via negative force constants.
\bibitem{siestacal2}
The same basis was used as in the infinite wire case \cite{siestacal}. 
This time the energy cutoff of the finite real-space grid was 100 Ry.
\end{references}
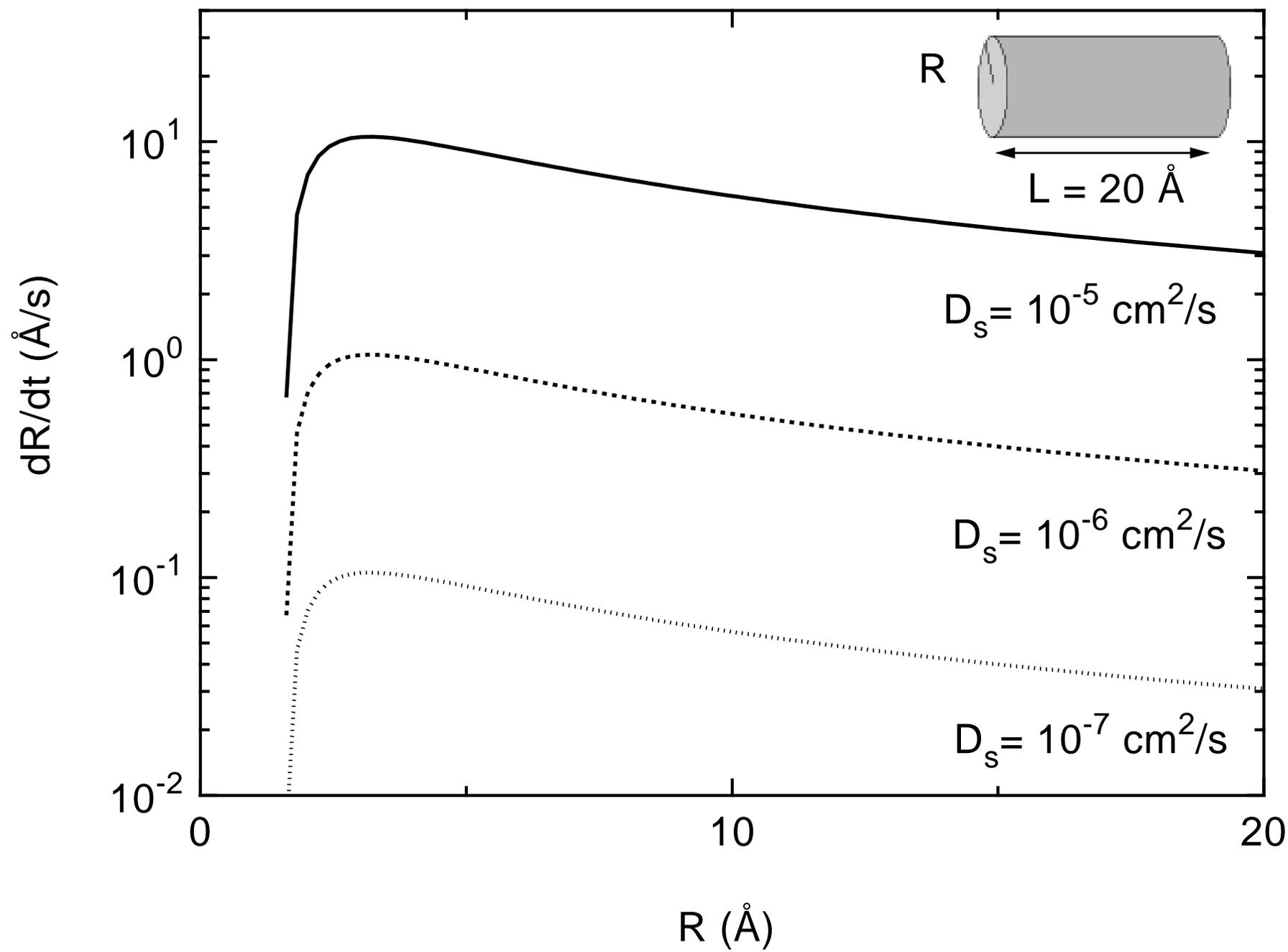
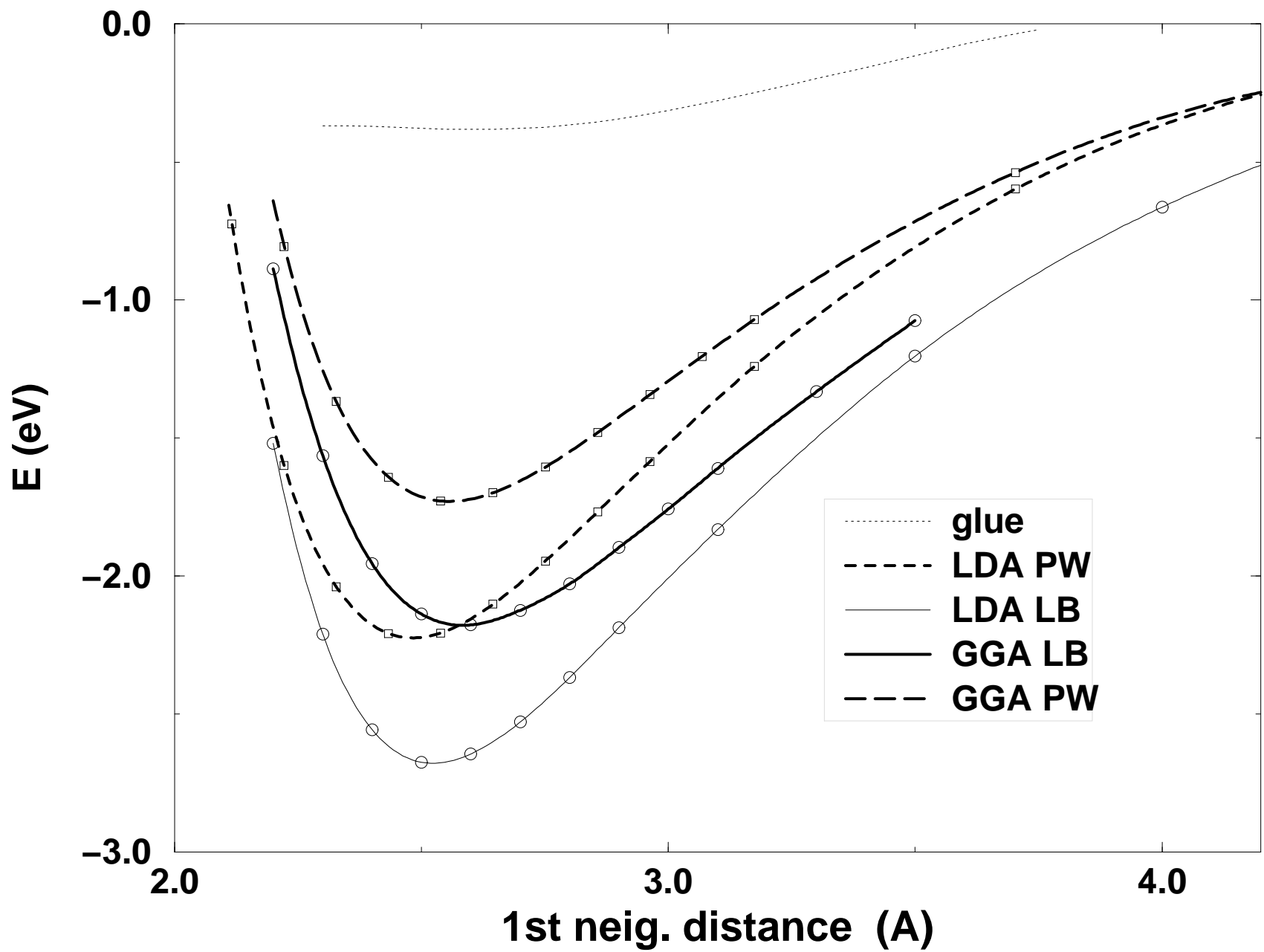
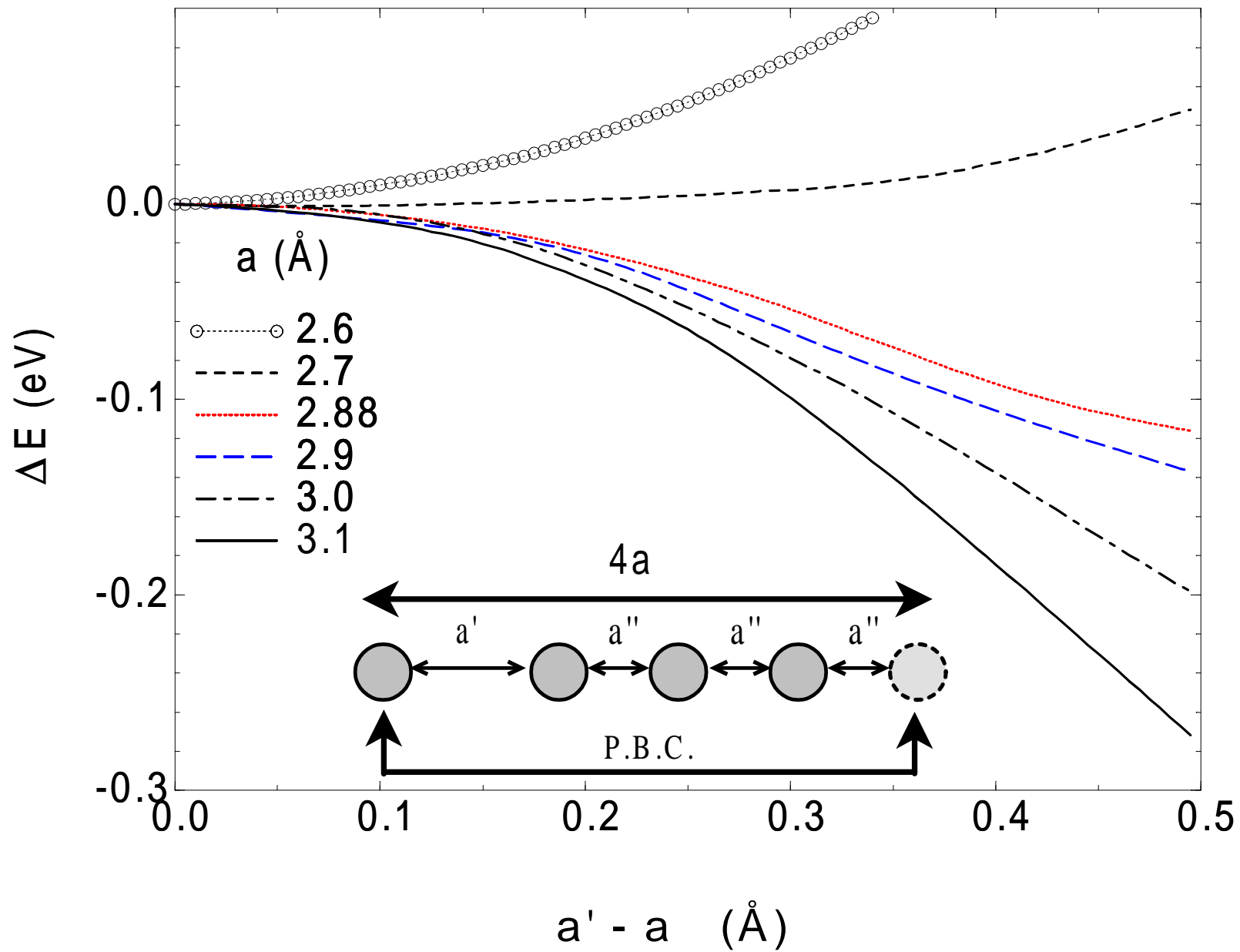
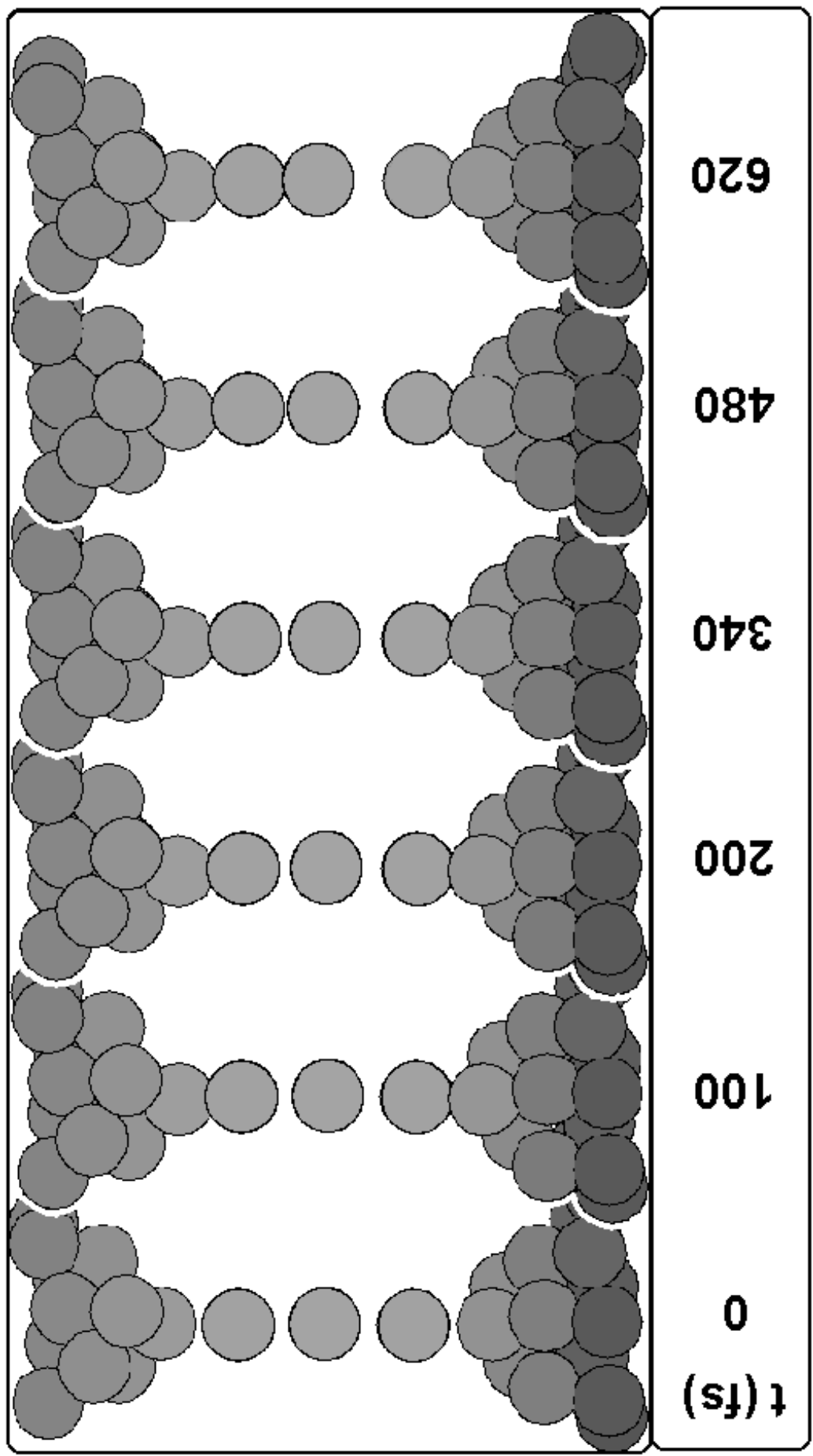
\end{document}